\catcode`\@=11
\def\gsim{\ifmmode{\mathrel{\mathpalette\@versim>}}
    \else{$\mathrel{\mathpalette\@versim>}$}\fi}
\def\lsim{\ifmmode{\mathrel{\mathpalette\@versim<}}
    \else{$\mathrel{\mathpalette\@versim<}$}\fi}
\def\@versim#1#2{\lower 2.9truept \vbox{\baselineskip 0pt \lineskip
    0.5truept \ialign{$\m@th#1\hfil##\hfil$\crcr#2\crcr\sim\crcr}}}
\catcode`\@=12
\def\msun{\hbox{$M_\odot$}}

\def\pn{\par\noindent}

\def\m*{\hbox{$M_*$}}

\def\ho{\hbox{$H_\circ$}}
\def\h50{\hbox{$\ho /50$}}
\def\h70{\hbox{$h_{70$}}}
\def\yr-1{\hbox{${\rm yr}^{-1}$}}

\documentclass[vecphys]{svmult}

\usepackage{graphicx}        
\usepackage{multicol}        
\usepackage[bottom]{footmisc}

\makeindex             

\begin{document}

\title*{Journeying the Redshift Desert}
\author{Alvio Renzini\inst{1} \& Emanuele Daddi\inst{2}}
\institute{INAF - Osservatorio Astronomico di Padova, Italy
\and CEA, Saclay, France}
%
%
\maketitle
\vskip -1.5cm
\begin{figure}
\includegraphics[height=11.8cm, width=3truecm, angle=-90]{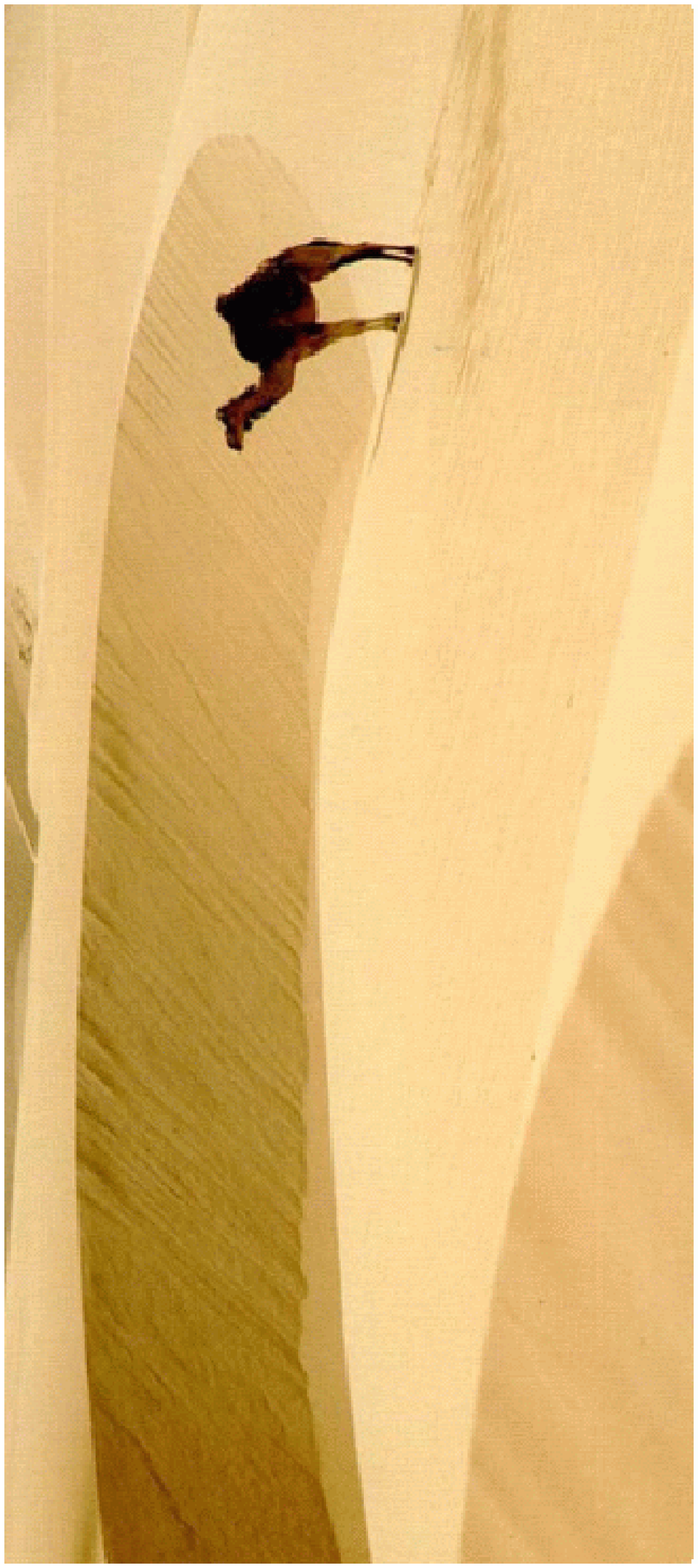}
\label{Fig. 0}
\end{figure}

{\bf The cosmic star formation rate, AGN activity, galaxy growth, mass
assembly and morphological differentiation all culminate at redshift
$\sim 2$. Yet, the redshift interval $1.4\lsim z\lsim 3$ is harder to
explore than the closer and the more distant Universe. In spite of so
much action taking place in this spacetime portion of the Universe, it
has been dubbed the {\it Redshift Desert}, as if very little was
happening within its boundaries. The difficulties encountered in
properly mapping the galaxy populations inhabiting the Desert are
illustrated in this paper, along with some possible remedy.}

\bigskip\noindent
{\bf Optical Spectroscopy of $1.4\lsim z\lsim 3$ galaxies}

\medskip\noindent
Fig. 1 shows typical FORS2 spectra of actively starforming, moderately
starforming, and passively evolving galaxies at $z\lsim 1$ (Mignoli et
al. 2005). The strongest, most easily recognizable features in these
spectra are the [OII]$\lambda$3727 line in emission, the CaII H\&K
doublet, and next to it the 4000 \AA\ break. These are the features
that allow spectroscopists to measure reliable redshifts even on
relatively low S/N spectra. Of course, provided they are included in
the observed spectral range.

As redshift increases beyond $z\sim 1$ all these features become
harder to recognize in observed spectra, as they enter a wavelength
region where the sensitivity of CCDs starts to drop, fringing
complicates the life, and sky deteriorates.  At this point we are already in a
quite arid environment (redshift-wise), though still manageable
thanks to the collective power of our very large telescopes, routinely
applied dithering patterns, and the like. But a little further out in redshift
our preferred spectral features move beyond 1 $\mu$m, i.e., into the near-IR,
and we are in full desert.

Still, survival with optical spectrographs is hard, but not completely
impossible.  While we have lost beyond 1 $\mu$m the strong features,
other, albeit less prominent ones have entered our optical range
coming from the rest frame ultraviolet. As opportunistic organisms
still find their ecological niche in a driest desert, so we currently
dwell on weak, rest-frame UV feature to explore the redshift
desert. In the case of actively starforming galaxies at $z\gsim 1.4$,
these are several narrow absorption lines over the UV continuum, most
of which originating in the ISM of these galaxies (see Fig. 2). In the
case of passively evolving {\it elliptical} galaxies at $z>1.4$, the
strongest feature in the observed optical spectral range is a
characteristic feature at $\lambda\lambda\sim 2600-2850$ \AA, due to
neutral and singly ionized magnesium and iron (see Fig.3). Thanks of
them we can survive in the desert, but it is not an easy life.

\begin{figure}
\includegraphics[width=9cm, angle=-90]{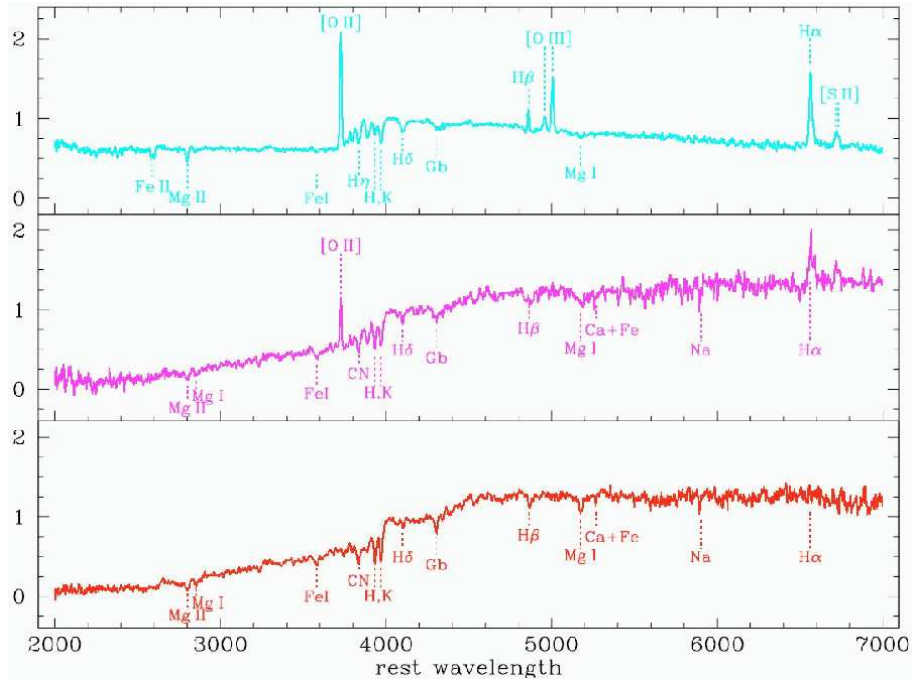} 
    \caption{The template spectra of actively starforming (top),
 moderately starforming (middle) and passively evolving galaxies (bottom)
(from Mignoli et al. 2005). 
}
\label{Fig. 1}
  \end{figure}

\begin{figure}
    \centering
\includegraphics[height=8cm, angle=-90]{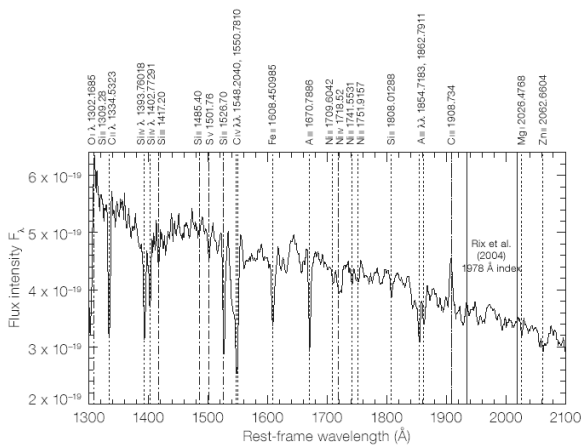} 
\caption{The co-added FORS2 spectrum of 75 starforming galaxies at $z\sim 2$,
corresponding to 1652.5 hours of integration (From Halliday et
al. 2008). The main spectral features are indicated, including the
weak blend of FeIII lines that originate in the photosphere of the OB
stars responsible for the UV continuum.}
 \label{Fig. 2}
\end{figure}

First of all, it is quite awkward to use narrow, weak absorption lines
to get redshifts of starforming galaxies that have strong emission
lines elsewhere in the spectrum, or absorptions on a very faint UV
continuum for galaxies that are intrinsically very red. These are
indeed the cases shown in Fig. 2 and 3! And this is not the whole
story.  To make the fairly good S/N spectra shown in these figures
from the GMASS Large Programme, Cimatti et al. (2008) had to coadd the
spectra of several galaxies, each integrated from a minimum of 30 to a
maximum of 60 hours. Thus, the spectrum of starforming galaxies in
Fig. 2 is the result of co-adding 75 spectra of individual galaxies
for a total integration time of 1652.5 hours (!), and the spectrum of
passive galaxies in Fig. 3 was obtained co-adding the spectra of 13
galaxies, for a total integration time of 480 hours (!).  Clearly,
journey the redshift desert nowadays takes time.

In the case of starforming galaxies a little relief may be offered by
Ly-$\alpha$, if the spectrograph is efficient enough in the
UV. Indeed, even if not in emission Ly-$\alpha$ is such a strong
feature that it helps a lot to get redshifts. However, in a
spectrograph such as e.g., VIMOS Ly-$\alpha$ does not get in before
$z\sim 1.8$, hence the range $1.4\lsim z \lsim 1.8$ is perhaps the
harshest part of the redshift desert.

\begin{figure}
    \centering
\includegraphics[height=8cm, angle=-90]{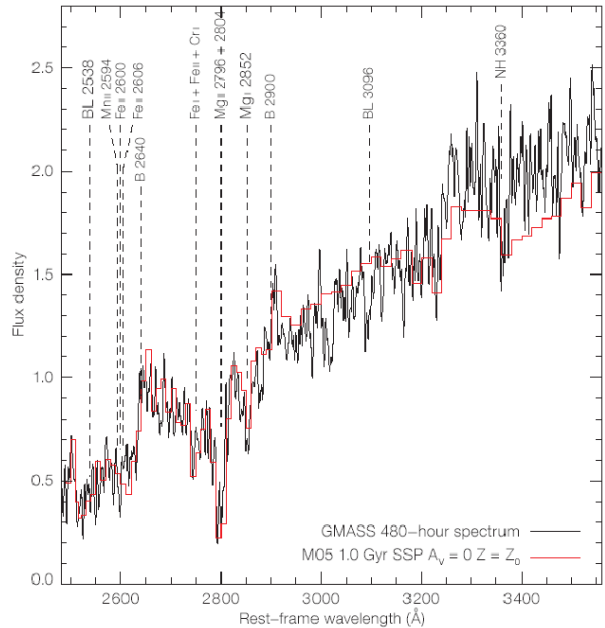} 
\caption{The co-added FORS2 spectrum of 13 passively evolving galaxies at 
$z\sim 1.6$, corresponding to 480 hours of integration (From Cimatti
et al. 2008). The main spectral features are indicated, along with the
synthetic spectrum of a 10 Gyr old, solar metallicity stellar
population model (Maraston et al. 2005).
}
 \label{Fig. 3}
\end{figure}

\medskip\noindent
{\bf Drawbacks } 
\medskip\noindent

Since quite a few years we know that at $z\sim 2$ galaxies with star
formation rates (SFR) as high as some $\sim 100\;\msun\yr-1$ are quite
common, and, by analogy with the rare objects at $z\simeq 0$ with
similar SFRs, many of us believed they were caught in a merging-driven
starburst. It was quite a surprise when one of these galaxies
(BzK-15504 at $z=2.38$) didn't show any sign of ongoing merging, but on
SINFONI 3D spectroscopy was looking as a rather ordered rotating disk
(Genzel et al. 2006). Still, with many clumps and high velocity dispersion
making it (like many others, see F\"orster-Schreiber et al. 2009)
quite different from local disk galaxies.

That high SFRs in $z\sim 2$ galaxies does not necessarily implies {\it
starburst} activity became clear from a study of galaxies in the GOODS
fields (Daddi et al. 2007a). Fig. 4 shows the SFR vs stellar mass $M*$
for galaxies at $1.4\lsim z\lsim 2.5$ in the GOODS-South field, where
a tight correlation is apparent between SFR and stellar mass. Only a
few galaxies are far away from the correlation, most notably a
relatively small number of passive galaxies (with undetectable SFR),
conventionally placed at the bottom of Fig. 4. Among starforming
galaxies, the small dispersion of the SFR for given $M*$ demonstrates
that these objects cannot have been caught in a special, starburst
moment of their existence.  Rather, they must sustain such high SFRs
for a major fraction of the time interval between $z=2.5$ and $z=1.4$,
i.e. for some $10^9$ yr instead of the order of one dynamical time
($\sim 10^8$ yr) typical of starbursts. 

In parallel with these observational evidences, theorists are shifting
their interests from (major) mergers as the main mechanism to grow
galaxies, to continuous {\it cold stream} accretion of baryons, hence
turned into stars (Deckel et al. 2009). Clearly a continuous, albeit
fluctuating SFR such as in these models is far more keen to the
evidence shown in Fig. 4, compared to a scenario in which star
formation proceeds through a series of short starbursts interleaved by
long periods of reduced activity. This is not to say that major mergers
don't play a role. They certainly exists, and can lead to real giant starbursts
bringing to SFRs as high as $\sim 1000\;\msun\yr-1$, currently identified with 
submillimeter galaxies (e.g. Tacconi et al. 2008).

\begin{figure}
    \centering
\includegraphics[height=8cm, angle=0]{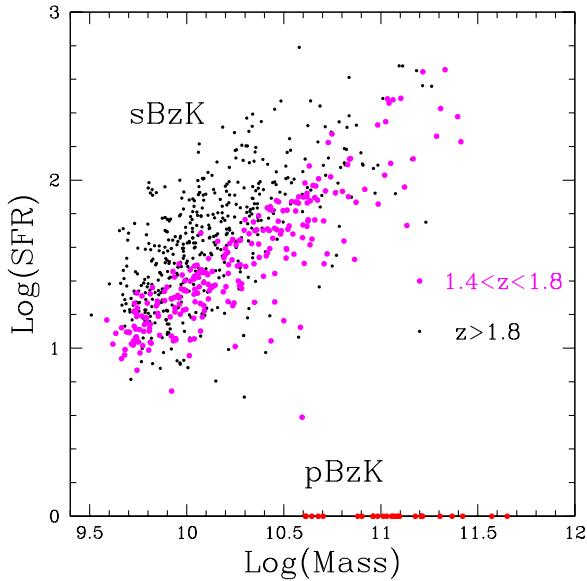} 
\caption{The SFR in $\msun\yr-1$ vs stellar mass for actively starforming 
galaxies galaxies (sBzK) in the GOODS-South field and with
spectroscopic or photometric redshifts in the range $1.4<z<2.5$
(adapted from Daddi et al. 2007a). Passively evolving galaxies (with
SFR $\simeq 0$, dubbed pBzKs from Daddi et al. 2004) are
conventionally plotted at the bottom as red dots.}
\label{Fig. 4}
\end{figure}

This {\it paradigm shift}, from mergers to cold streams, adds flavor
to a thoroughly exploration of the redshift desert, an enterprise
which is at the core of the zCOSMOS-Deep project (Lilly et al. 2007),
the largest ongoing spectroscopic survey of the desert. This survey is
targeting starforming galaxies whose spectrum is pretty much like that
shown in Fig. 2, and does so with VIMOS for objects down to $B$
magnitude $\sim 25$ with 5 hours integrations. The success rate of
zCOSMOS-Deep (i.e., the fraction of targets for which a reliable
redshift is obtained) is $\sim 2/3$ (Lilly et al. in preparation), not
bad at all for objects in the desert! Still, we wonder what we get,
and what we miss.

\begin{figure}
    \centering
\includegraphics[height=8cm, angle=0]{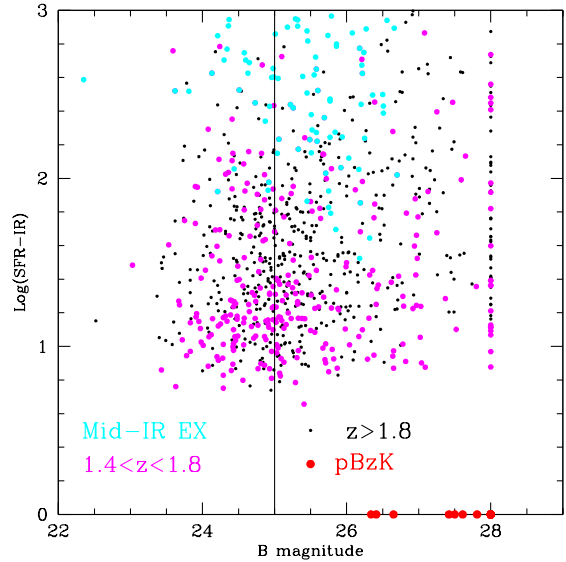} 
\caption{The SFR (here measured from the 24 $\mu$m flux as in Daddi et al. 
2007a) vs the $B$ magnitude for $1.4<z<2.5$ galaxies in the
GOODS-South field. The cyan dots denote galaxies with excess mid-IR
emission, which according to Daddi et al. (2007b) may be due to a
buried, Compton-thick AGN, in which case the SFR may have been
overestimated. The vertical line marks the current practical limit of 
what is doable with the VIMOS instrument.}
\label{Fig. 5}
\end{figure}

Fig. 5 shows the SFR vs $B$ magnitude for the same $1.4\lsim z\lsim
2.5$ GOODS galaxies shown in Fig. 4. Clearly, the vast majority of
actively starforming galaxies in the desert are fainter than $B=25$,
and they include several among the most active galaxies (here and
elsewhere magnitudes are in the AB system). Those brighter than $B=25$
account for just $\sim 16\%$ of the global SFR of the whole sample,
hence $\sim 84\%$ of it remains out of reach. But why is the $B$
magnitude (i.e., the rest-frame UV) such a poor indicator of SFR? This
is so because to sustain high SFRs one needs lots of gas, gas is
accompanied by dust, and dust is a potent absorber of UV radiation.

Fig. 6 shows the dust reddening $E(B-V)$ for the same set of GOODS
galaxies, as a function of SFR (from Greggio et al. 2008). Indeed, the
most starforming galaxies are also the most extincted ones, which
makes it difficult to get redshifts from blue-band spectroscopy.  Such
a strong correlation of extinction and SFR has been recently
quantitatively confirmed using the dust-free 1.4 GHz flux as a SFR
indicator (Pannella et al. 2009). What said for the SFR holds true
also for the stellar mass. Fig. 7 shows $M*$ vs $B$ magnitude, and
again most of the stellar mass is in galaxies fainter than $B=25$,
including many among the most massive galaxies.

\begin{figure}
    \centering
\includegraphics[height=8cm, angle=0]{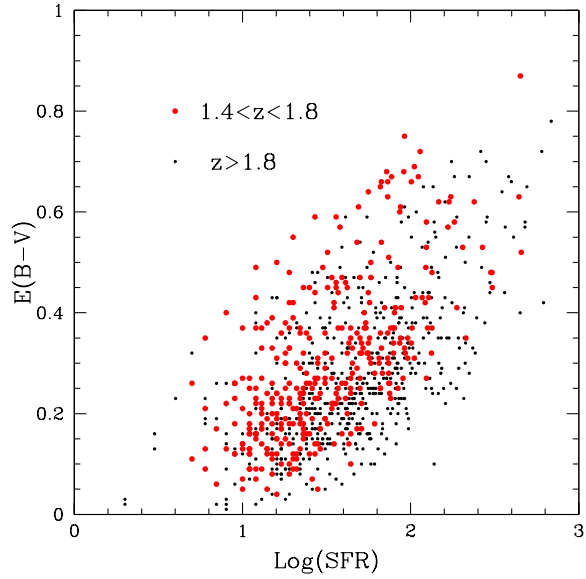}
\caption{The global reddening $E(B-V)$ derived from the slope of the 
rest-frame UV continuum as a function of the star formation rate for
the same objects shown in Fig. 5.  }
\label{Fig. 6}
\end{figure}

\begin{figure}
    \centering
\includegraphics[height=8cm, angle=0]{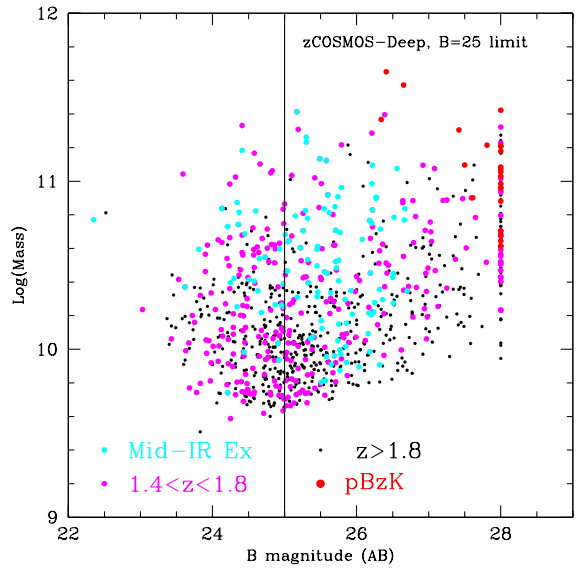} 
\caption{The stellar mass vs  $B$ magnitude for the same objects shown 
in Fig. 5. 
}
\label{Fig. 7}
\end{figure}

\begin{figure}
    \centering
\includegraphics[height=8cm, angle=0]{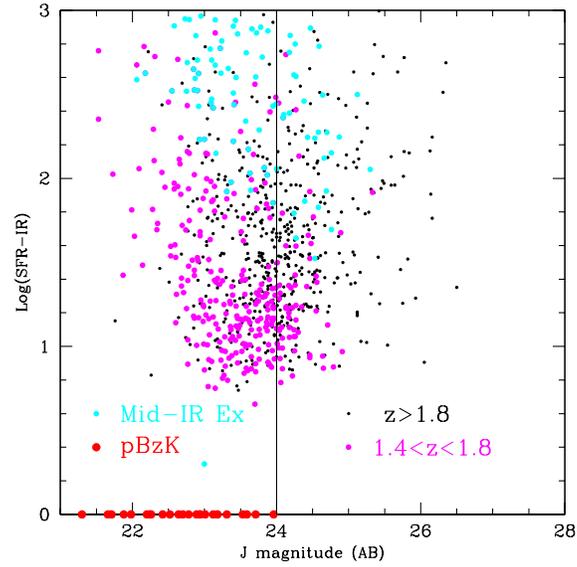}
\caption{The same as in Fig. 5, but now plotted vs the $J$ magnitude.
The vertical line at $J$(AB)=24 marks the limit expected for reaching
S/N=5 with 10h integration with the FMOS $J$-band spectrograph at the
SUBARU Telescope.
 }
\label{Fig. 8}
\end{figure}

\begin{figure}
    \centering
\includegraphics[height=8cm, angle=0]{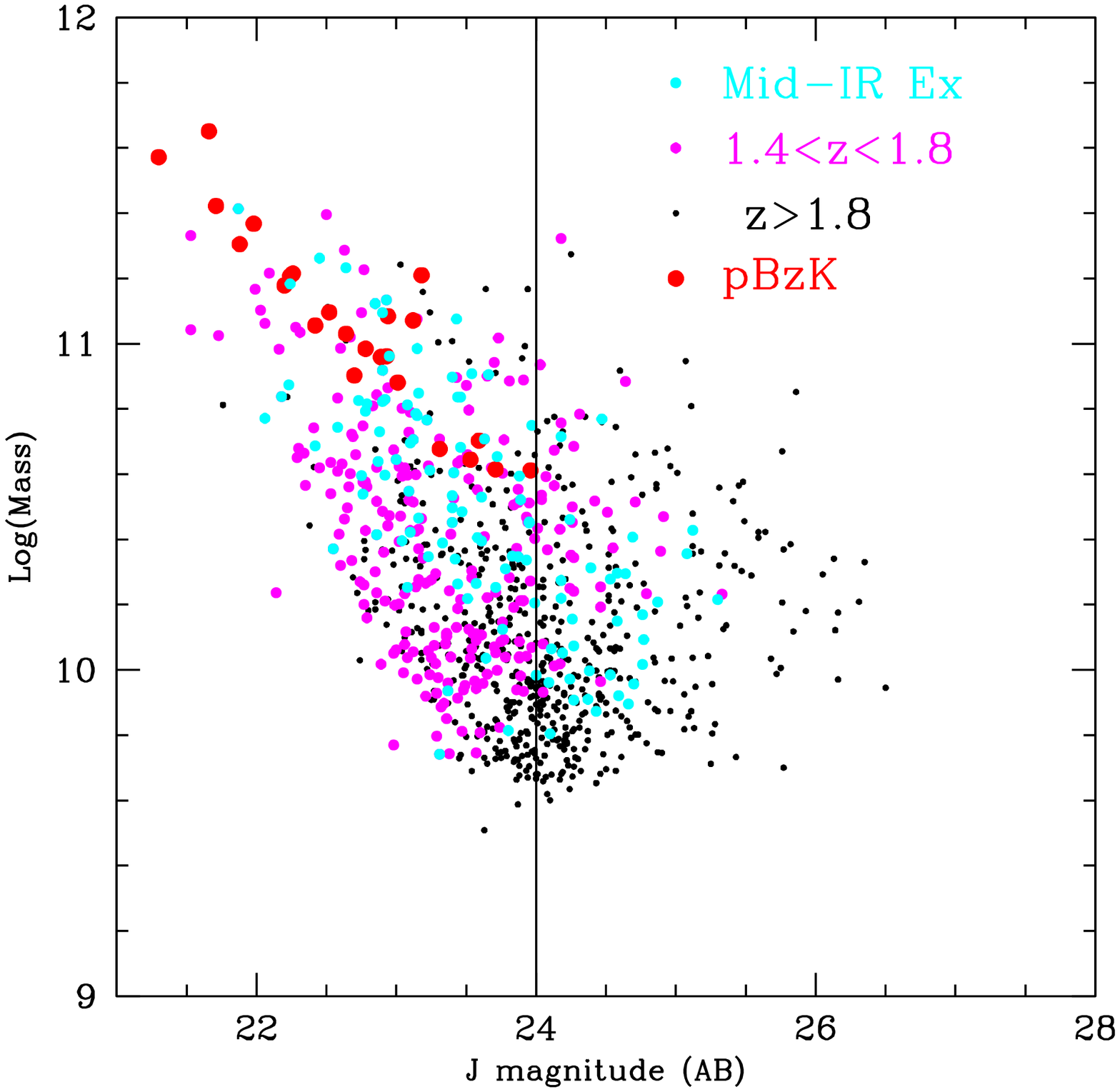}
\caption{The same as in Fig. 7, but now plotted vs the $J$ magnitude.
 }
\label{Fig. 9}
\end{figure}

Usually, when extinction bothers us it helps going in the near-IR.
Fig. 8 and and Fig. 9 are analogous to the previous two figures, but
SFR and $M*$ are now plotted vs the $J$-band magnitude instead of the
$B$ band.  Clearly, whereas a $B<25$ selection misses most of the SFR
and most of the stellar mass at $z\sim 2$, a $J<24$ selection would
pick most of them.  In particular, note that most of the most
starforming and most massive galaxies are fainter than $B=25$, they
are instead among the brightest in the $J$ band. Thus, a $B<25$
selection picks a fair number of massive, starforming galaxies at
$z\sim 2$, but misses the majority of them, and in particular may miss
several of the most massive and most starforming ones.

It is worth emphasizing that a comparison of Fig. 7, 8 and Fig. 9
shows that all passive galaxies (the pBzKs of Daddi et al. 2004) are
among the faintest objects in the $B$ band, but are among the
brightest ones in the $J$ band. Being fainter than $B=25$, all passive
galaxies at $z>1.4$ are automatically excluded from e.g., the zCOSMOS
survey. Now, there are over 3,000 such galaxies in the COSMOS field
(McCracken et al 2009), and if we wanted to make on the whole COSMOS
field (7200 arcmin$^2$) the same effort that GMASS did on one FORS2
FoV (49 arcmin$^2$) investing over 100 hours of VLT time, then it
would take well over 15,000 hours (!)  of telescope time.  Passive
galaxies at $z>1.4$ are the most massive galaxies at these redshifts,
and they likely mark the highest density peaks in the large scale
structure, but we suspect that this would not be sufficient for the OPC
to recommend the allocation of over 1,500 VLT nights to such a project...

Having touched upon the COSMOS field, using COSMOS data for 30,866
starforming galaxies Fig. 10 and Fig. 11 further illustrate the
differences between $B$-band and $J$-band limited samples of $z\sim 2$
galaxies. Galaxies are first selected with the BzK criterion of Daddi
et al. (2004) from the COSMOS $K$-band catalog (McCracken et
al. 2009), which is complete down to $K=23.5$. Then multiband
photometric redshift from Ilbert et al. (2009) are used.  Notice that
the full range of masses and SFRs are still sampled for a selection
down to a limit magnitude as bright as $J=22-23$.  In Fig.s 8-11 the
vertical line at $J=24$ is meant for objects that would be detected
with S/N=5 with 10h integrations with the FMOS $J$-band spectrograph at 
the SUBARU telescope (Kimura et al. 2003). This may well be a rather optimistic
limit for a robust detection of the continuum and the absorption lines 
of passive galaxies. But for starforming galaxies the [OII] emission line
would much help in measuring redshifts, hence a $J=24$ limit may not be 
mere dream for such objects.  

\begin{figure}[ht]
    \centering
\includegraphics[width=7cm, angle=-90]{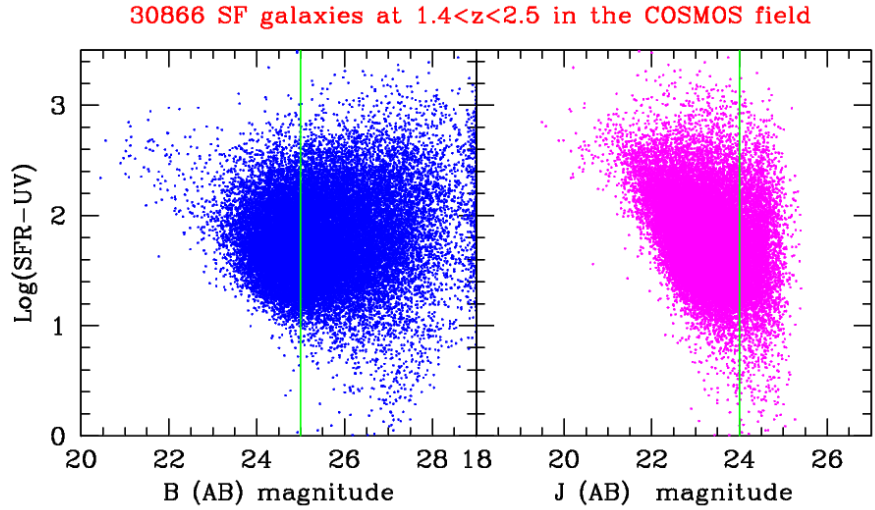}
\caption{The SFR from the UV diagnostics for SF galaxies at 
$1.4 <z<2.5$ in the COSMOS field vs. their $B$ magnitude (left) 
and their $J$ magnitude (right). A plume of objects brighter than $B\sim 22$ 
are likely to be AGN.}
\label{Fig. 10}
\end{figure}

\begin{figure}[b]
    \centering
\includegraphics[width=7cm, angle=-90]{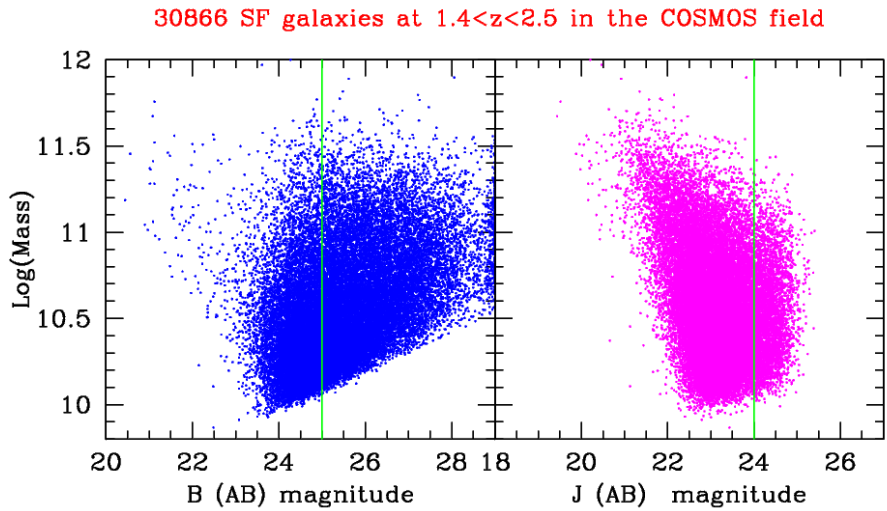}
\caption{The same as in Fig. 10, but now plotted is the stellar mass vs the 
$B$ and $J$ magnitudes.
 }
\label{Fig. 11}
\end{figure}

\medskip 
\noindent
{\bf Remedies}

\medskip 
\noindent
We understand that many may prefer to leave deserts as uncontaminated
as possible, rather than crowded by swarms of all-inclusive
tourists. But, what options do we have if we really want to fully
colonize the redshift desert?

One possibility would be to use VIMOS with much longer integrations
compared to the 5h currently invested by the zCOSMOS project, i.e.,
$\gsim 30$h as used for the GMASS project. But before doing so VIMOS
would have to be made at least as efficient as FORS2 in the red, a
good thing that may happen anyway. FoV-wise, VIMOS is like 4 FORSes,
hence doing all the COSMOS pBzKs (and along with them a much larger
number of starforming galaxies in the desert) would take $\sim 1/4$ of
the time we have estimated above for FORS2, i.e., some 350 VLT
nights. This still looks a lot of time, yet somewhat more affordable
than a mere FORS2 brute force effort. After all VIMOS was conceived
and built for making primarily large redshift surveys, hence, why not
this one? But, ``how many years are 350 nights?'' We can scale from
zCOSMOS, whose 640 hours ($\sim 75$ nights) were calibrated to
complete the project in 4 semesters. If (big if) VIMOS could be used
for zCOSMOS whenever the COSMOS field is $\pm 4$ hours from the
meridian. But because of bad weather, projects competing for objects
at the same r.a., and instrument downtime, it is now taking 5 years to
finish zCOSMOS.  By the same token, it would then take $\sim 25$ years
to an upgraded VIMOS to do justice of just the COSMOS field.

Thus, what we would really need is a high-multiplex instrument able to
sample the strongest spectral features of galaxies in the $1.4<z<2.5$
desert, i.e., [OII]3727 for the overwhelming population of SF galaxies, 
and CaII H\&K and the 4000 \AA\ break for the passive ones. All these
features fall in the $J$ band for the galaxies in the desert, thus a
cryogenic instrument would not be necessary. Without having to bother 
for the thermal background, a room temperature instrument could then 
cover wide fields in a single telescope pointing. 
A preliminary knowledge of the distribution of the [OII]-line flux for
starforming galaxies in the desert would be critical for properly planning 
a spectroscopic survey targeting them. Such information is not yet in hand.

The surface density of these objects for the full COSMOS sample down
to $K\sim 23.5$ is $\sim 4/{\rm arcmin}^2$, or $\sim 1/{\rm arcmin}^2$
for their brighter portion down to $J=22$. Thus, the ideal VLT
instrument would be one able to fully exploit the largest FoV of the
VLT (i.e., $\sim 500$ arcmin$^2$ at the Nasmyth) with a multiplex
$\gsim 1$ arcmin$^{-2}$, or $\sim 500$ over the whole field. This can
be achieved only with a fiber-fed $zJ$-band spectrograph, not too
different from the FMOS instrument on SUBARU.  A {\it camel} of this
species may offer the best, short-term possibility of journeying the
redshift desert.





\begin{thebibliography}{99.}

\bibitem{} Cimatti, A. et al. 2008, A\&A, 482, 21
\bibitem{} Daddi, E. et al. 2004, ApJ, 617, 746
\bibitem{} Daddi, E. et al. 2007a, ApJ, 670, 156
\bibitem{} Daddi, E. et al. 2007b, ApJ, 670, 173
\bibitem{} Deckel, A. et al. 2009, Nature, 457, 451
\bibitem{} F\"oster-Schreiber, N.M. et al.  2009,  ApJ, submitted 
           (arXiv0903.1872)
\bibitem{} Genzel, R. et al. 2006, Nature, 442, 786
\bibitem{} Greggio, L. et al. 2008, MNRAS, 388, 829
\bibitem{} Halliday, C. et al. 2008, A\&A, 479, 417
\bibitem{} Kimura, M., et al. 2003, SPIE, 4841, 974
\bibitem{} Lilly, S.J. et al. 2007, ApJS, 172, 70
\bibitem{} Maraston, C. 2005, MNRAS, 362, 799
\bibitem{} McCracken, H.J. et al. 2009, ApJ, submitted
\bibitem{} Mignoli, M. et al. 2005, A\&A, 437, 883
\bibitem{} Pannella, M., et al. 2009, ApJ, 698, L116
\bibitem{} Tacconi, L.J. et al. 2008, ApJ, 680, 246
\end{thebibliography}
\end{document}